\documentclass[aps,prl,reprint,groupedaddress]{revtex4-1}
\usepackage[utf8]{inputenc}
\usepackage{siunitx}
\usepackage{placeins}
\usepackage{color}
\usepackage{amssymb}
\usepackage{amsmath}
\usepackage{miller}
\usepackage[english]{babel}
\usepackage{graphicx}
\usepackage{color}

\newcommand{\cts}{cts}

\newcommand{\ket}[1]{{\left|#1\right>}}

\newcommand{\seconds}{\second}
\newcommand{\citein}[1]{Ref.~\onlinecite{#1}}
\newcommand{\textsubscript}[1]{$_{\text{#1}}$}
\newcommand{\nv}{NV\textsuperscript{-}}
\newcommand{\nvn}{NV\textsuperscript{0}}
\newcommand{\mean}[1]{\overline{#1}}
\newcommand{\dt}{\mathrm{d}t}
\newcommand{\nsn}{$\text{N}_\text{s}^0$}
\newcommand{\nsnm}{\text{N}_\text{s}^0}

\DeclareSIUnit{\sqrthz}{\ensuremath{\sqrt{\text{\hertz}}}}

\begin{document}
\title{Efficient electrical spin readout of  \nv~centers in diamond}

\author{Florian M. Hrubesch}
\email{florian.hrubesch@wsi.tum.de}
\affiliation{Walter Schottky Institut and Physik-Department, Technische Universität München, Am Coulombwall 4, 
85748 Garching, Germany}

\author{Georg Braunbeck}
\affiliation{Walter Schottky Institut and Physik-Department, Technische Universität München, Am Coulombwall 4, 
	85748 Garching, Germany}

\author{Martin Stutzmann}
\affiliation{Walter Schottky Institut and Physik-Department, Technische Universität München, Am Coulombwall 4, 
	85748 Garching, Germany}

\author{Friedemann Reinhard}
\email{friedemann.reinhard@wsi.tum.de}
\affiliation{Walter Schottky Institut and Physik-Department, Technische Universität München, Am Coulombwall 4, 
	85748 Garching, Germany}

\author{Martin S. Brandt}
\email{brandt@wsi.tum.de}
\affiliation{Walter Schottky Institut and Physik-Department, Technische Universität München, Am Coulombwall 4, 
	85748 Garching, Germany}

\begin{abstract}
Using pulsed photoionization the coherent spin manipulation and echo formation of ensembles of \nv~centers in diamond are detected electrically realizing contrasts of up to \SI{17}{\percent}. The underlying spin-dependent ionization dynamics are investigated experimentally and compared to Monte-Carlo simulations. This allows the identification of the conditions optimizing contrast and sensitivity which compare favorably with respect to optical detection.
\end{abstract}

\maketitle

The nitrogen vacancy center \nv~in diamond is a promising candidate for quantum applications, since its coherence time at room temperature is in the range of ms \cite{Bar-Gill2013} and its spin can be read out by optical fluorescence detection \cite{Gruber1997}. These features have enabled the use of \nv~centers, e.g.,~as a quantum sensor for magnetic fields \cite{taylor2008, Wolf2015} and temperature \cite{Acosta2010}
, for scanning-probe spin imaging \cite{maletinsky12} and structure determination of single biological molecules \cite{Shi2015}. Despite its apparent simplicity, however, optical spin readout has drawbacks: it is highly inefficient, requiring several $100$ repetitions for a single spin readout, and cumbersome to implement in many applications. Electric readout of spin in a suitable diamond semiconductor device appears as an attractive way to surmount these limitations. It could enable access to \nv~centers in dense arrays, with a spacing limited by the few-nm-small feature size of electron beam lithography \cite{Manfrinato2013} rather than the optical wavelength. It might, moreover, provide a way to read out other spin defects \cite{Jungwirth2014, christle2015,Widmann2015}, potentially including optically inactive ones. 

Two methods for electric readout of \nv~centers have been demonstrated. The method in \citein{Brenneis2015} uses non-radiative energy transfer to graphene and detects the spin signal in the current through the graphene sheet generated by this transfer. In contrast, the method presented in \citein{Bourgeois2015} uses the charge carriers generated directly in the diamond host crystal by photoionization of the \nv~centers (photocurrent detection of magnetic resonance, PDMR). Both methods, however, have until now only been used with continuous wave (cw) spin manipulation and have therefore remained limited to \nv~detection. Here we demonstrate a scheme based on both pulsed spin manipulation and pulsed photoionization to truely read out the spin state of \nv~centers electrically after coherent control, using Rabi oscillations and echo experiments as examples. We employ this scheme to establish a quantitative model of photoionization,  simulate the readout efficiency and predict, that under optimized conditions pulsed electric readout could outperform optical fluorescence detection.

\begin{figure}[t]
\centering
\includegraphics[width=\linewidth]{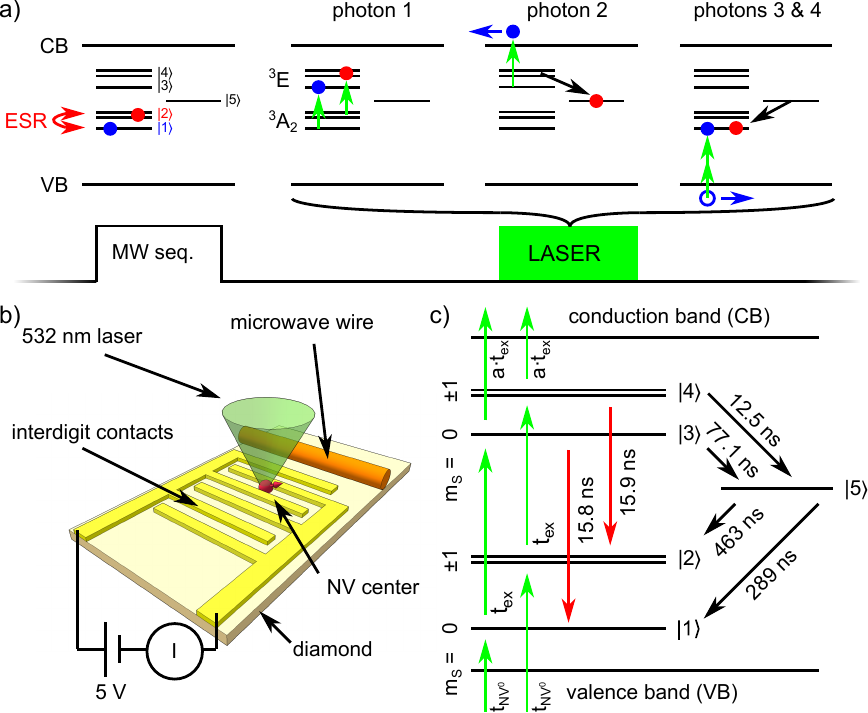}
\caption{a)
Spin-dependent photoionization of \nv~centers used for the electrical readout of its spin state. b) Schematic drawing of the sample and the measurement setup. c)  Level scheme used for the Monte-Carlo simulation.}
 \label{process}
\end{figure}

The spin-dependent photoionization cycle can be understood as an effective four-photon process, whose spin dependence relies on the \nv~center's inter-system crossing (ISC) which is also key to the classic optical readout (Fig.~\ref{process} a)) \cite{Bourgeois2015}. A first photon (green arrows) triggers shelving (black arrow) of \nv~centers in spin state $\ket 2$ (corresponding to the $m_\text{S}=\pm 1$ spin quantum numbers of \nv) into the long-lived metastable singlet state $\ket 5$ by this ISC. Since shelving protects this spin state from further laser excitation, absorption of a second photon preferentially ionizes \nv~centers prepared in spin state $\ket 1$ (corresponding to $m_\text{S}=0$) into the conduction band (CB), creating a spin-dependent photocurrent (blue arrow) proportional to the population of the $m_\text{S}=0$ state obtained by a microwave pulse sequence (red arrow) preceding the optical pulse. Two further photons re-charge the \nvn~center into its negative charge state by exciting the \nvn~(photon 3) and capture of an electron (photon 4) from the valence band (VB) \cite{Siyushev2013}. 

Our spin readout experiments are performed in a photoconductor as shown in Fig.~\ref{process} b). We illuminate a densely \nv-doped diamond (Element 6, grown by chemical vapor deposition, with [N] $<1$~ppm, [NV] $\approx 10$~ppb) with a green laser (wavelength \SI{532}{\nano\meter}) pulse generated by a Nd:YAG laser and an acousto-optic modulator (AOM) and observe the resulting photocurrent between two interdigit Schottky contacts, biased with a voltage of \SI{5}{\volt}.  
The contacts (finger width \SI{5}{\micro\meter}, finger-to-finger distance \SI{10}{\micro\meter}) consist of a 10-nm-thick titanium layer and a 80-nm-thick gold layer, deposited on the diamond surface after cleaning it in a H\textsubscript{2}SO\textsubscript{4}/H\textsubscript{2}O\textsubscript{2} mixture, followed by an oxygen plasma treatment. The photocurrent through the sample is measured using a transimpedance amplifier (amplification \SI{1}{\giga\volt\per\ampere}, bandwidth \SI{10}{\hertz}). 
Depending on the measurement we use a \num{5}x, a \num{10}x or a \num{100}x objective, with numeric apertures of \num{0.15}, \num{0.30} and \num{0.80} and diffraction limited spot sizes of \SI{3500}{\nano\meter}, \SI{1800}{\nano\meter} and \SI{670}{\nano\meter}, resp.The microwave with frequency $\nu_\text{MW}$ for spin manipulation is delivered to the sample using a wire next to the interdigit contact structure (cf. Fig.~\ref{process} b)).

\begin{figure}[t]
\centering
\includegraphics[width=\linewidth]{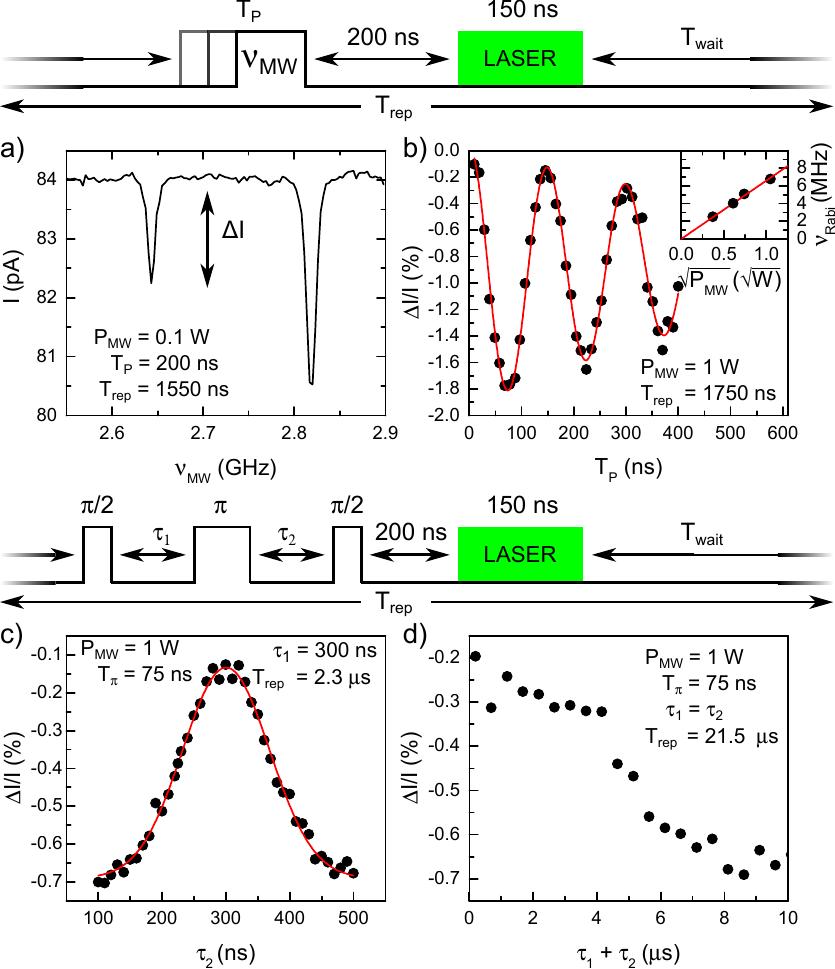}
\caption{a) Pulsed electrically detected magnetic resonance spectrum for $B_0\parallel\left<111\right>.$ b) Rabi oscillations in the contrast $\Delta I/I$ (symbols) with fit of an exponentially decaying cosine (line). The inset shows the frequency of Rabi oscillations (symbols) measured at different microwave powers and a linear fit (line). c) Spin echo measurement (symbols) with fit of a Gaussian (line). d) Echo decay measurement showing the first ESEEM decay. The pulse sequences used for the measurements are shown on top of the respective figure. $T_{\pi}$ is the length of a $\pi$ pulse.}
 \label{pEDMR}
\end{figure}

We first demonstrate that coherent control of the \nv~centers can be detected electrically using the pulse sequence shown on top of Fig.~\ref{pEDMR} a) and b). To excite electron spin resonance (ESR) transitions, the sequence starts with a microwave pulse with power $P_\text{MW}$ and varying duration $T_\text{P}$. This initializes the spin of the \nv~\textsuperscript{3}A\textsubscript{2} ground state. After a brief delay an optical excitation pulse follows (\num{10}x objective, light power of \SI{210}{\milli\watt} during the optical pulse). Furthermore, an external magnetic field of $B_0=\SI{8.1}{\milli\tesla}$ is applied to the sample parallel to one of the \hkl{111} axes via a permanent magnet so that only one crystallographic \nv~direction can be addressed. Figure \ref{pEDMR} a) shows the pulsed electrically detected magnetic resonance (pEDMR) spectrum obtained under these conditions, monitoring the dc current through the interdigit contact structure. In contrast to previous pEDMR experiments on silicon and organic semiconductors, where the spin dependence of comparatively slow recombination or hopping processes is monitored via a boxcar integration of the current transients following the spin manipulation \cite{Boehme2003, Stegner2006, Harneit2007, Hoehne2013, Kupijai2015}, the much faster pulse sequence repetition possible due to the fast photoioniziation and spin state initialization allows this vastly simpler direct detection of the spin signal in the dc current. As an example, the pulse repetition time $T_\text{rep}$ is \SI{1.5}{\micro\seconds} in Fig.~\ref{pEDMR} a). On a background photocurrent level of $I=\SI{84}{\pico\ampere}$ resonant decreases of the photocurrent are observed at $\nu_\text{MW}=\SI{2.643}{\giga\hertz}$ and $\SI{2.818}{\giga\hertz}$, corresponding to one \hkl{111} orientation parallel to the $B_0$ field and three off-axis \hkl{111} orientations, resp. The resonant change of the current of $\Delta I=-\SI{1.5}{\pico\ampere}$ at \SI{2.643}{\giga\hertz} corresponds to a relative spin-dependent current change (contrast) of $\Delta I/I = -\SI{1.8}{\percent}$. 

Rabi oscillations are observed when the length $T_\text{P}$ of the microwave pulse is changed, adjusting the waiting time $T_\text{wait}$ to keep $T_\text{rep}$ constant at $\SI{1750}{\nano\second}$. Figure \ref{pEDMR} b) shows the expected oscillatory dependence of $\Delta I/I$ on $T_\text{P}$. That indeed Rabi oscillations are obtained is demonstrated in the inset of Fig.~\ref{pEDMR} b), where the characteristic linear dependence of the oscillation frequency $\nu_\text{Rabi}$ on $\sqrt{P_\text{MW}}$ and, therefore, on the microwave magnetic field $B_1$ is observed. The Rabi oscillations exhibit an effective dephasing time of \SI{600}{\nano\second}, in accordance with other results on diamond with neutral isotope composition \cite{Parker2015,Mizuochi2009}. In all experiments represented in Fig.~\ref{pEDMR} b) to d) $\Delta I/I$ was determined by cycling the microwave frequency between the resonant $\nu_\text{MW}=\SI{2.643}{\giga\hertz}$ and two nonresonant frequencies $\SI{2.61}{\giga\hertz}$ and $\SI{2.68}{\giga\hertz}$, resulting in a low-frequency lock-in amplification \cite{Hoehne2012}.

The pulsed electrical detection scheme developed here also allows to detect spin echos, e.g.,~by using the pulse sequence depicted on top of Fig.~\ref{pEDMR} c) and d). As in the case of optically detected magnetic resonance (ODMR) \cite{Breiland1973, Childress2006} and other pEDMR \cite{Huebl2008} experiments, the corresponding Hahn echo sequence needs to be extended by a final $\pi/2$ pulse, which projects the coherence echo to a polarisation accessible to optical or electrical readout. Figure \ref{pEDMR} c) shows the echo in the contrast $\Delta I/I$ as a function of $\tau_2$ for a fixed $\tau_1=\SI{300}{\nano\second}$. At $\tau_1=\tau_2$ the total microwave pulse applied equals a nutation of $2\pi$, so that the contrast is minimal, in full agreement with Fig.~\ref{pEDMR} b). For $\tau_2$ significantly smaller or longer than $\tau_1$, no coherence echo is formed and the final $\pi/2$ projection pulse leads to an equal distribution of spin states which favor or which do not favor photoionization \cite{Huebl2008,Franke2014}. Indeed, a maximum $\Delta I/I$ of \SI{-0.7}{\percent} is observed for $\tau_1\ll\tau_2$ or $\tau_1\gg\tau_2$, in reasonable agreement with the contrast for $\pi/2$ pulses found in the Rabi oscillation experiment.

Finally, these echo experiments can also be performed as a function of total evolution time $\tau_1+\tau_2$ with $\tau_1=\tau_2$, giving access, e.g.,~to decoherence and to weak hyperfine interaction via electron spin echo envelope modulation (ESEEM). Figure~\ref{pEDMR} d) shows such an echo decay experiment on the \SI{2.643}{\giga\hertz} resonance where the decay is caused by ESEEM \cite{Stanwix2010}. Again, the pulse sequence repetition time is kept constant, the long times $\tau_1+\tau_2$ necessary for this experiment reduce the signal-to-noise ratio. Nevertheless, the experiments summarized in Fig.~\ref{pEDMR} clearly demonstrate that all fundamental coherent experiments can be performed on the \nv~center with the electrical readout scheme developed here.

\begin{figure}[t]
\centering
\includegraphics[width=\linewidth]{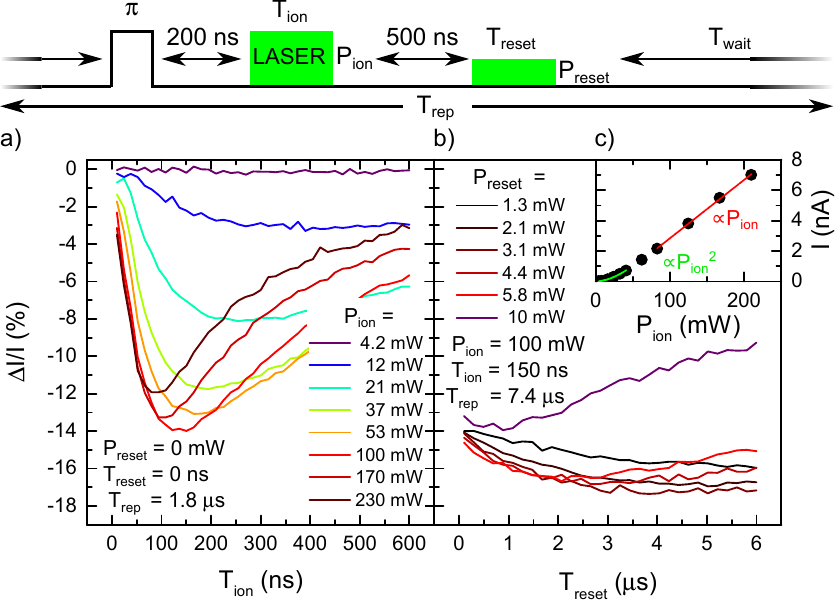}
\caption{a) Contrast $\Delta I/I$ for a $\pi$ microwave pulse as a function of the pulse length $T_\text{ion}$ of the ionization pulse for different ionization pulse powers $P_\text{ion}$. b) $\Delta I/I$ as a function of the pulse length $T_\text{reset}$ of an additional reset pulse for different reset pulse powers $P_\text{reset}$. The pulse sequence used is shown on top of the  figure. c) Photocurrent through the sample as a function of the laser power. The green line is a fit of a polynomial of degree two, the red line is a fit of a linear function.}
\label{tion}
\end{figure}
We now turn to study the readout contrast that can be obtained by pulsed electric readout. We therefore place ourselves at $B_0=0$, where all four \nv~orientations merge into a single resonance at $\nu_\text{MW}=\SI{2.87}{\giga\hertz}$, and where we can essentially flip \nv~with all orientations into state $\ket 2$ by a microwave $\pi$ pulse of \SI{75}{\nano\second} length using $P_\text{MW}=\SI{1}{\watt}$. Under these conditions we study the readout contrast as a function of both the duration $T_\text{ion}$ and intensity $P_\text{ion}$ of the readout light pulse (Fig.~\ref{tion} top), keeping illumination as homogeneous as possible by widening the laser beam with the \num{5}x objective to a spot size of \SI{3800}{\nano\meter}.

Figure \ref{tion} a) plots $\Delta I / I$ versus $T_\text{ion}$ for different $P_\text{ion}$. For each power we find an optimum pulse length between a regime of too short pulses, where the ionization of \nv~mostly takes place on a timescale faster than the shelving process suppressing the spin selection via the ISC, and too long pulses, where mostly \nv~contribute to the current which have lost their spin information by a decay through the ISC or by a preceding ionization. Optimizing $P_\text{ion}$ and $T_\text{ion}$ for the sample studied, we reach an optimum contrast of $\SI{-14}{\percent}$ at an intermediate power of \SI{100}{\milli\watt}. As will be discussed below, this value is probably limited by ionization of background substitutional nitrogen donors (\nsn) \cite{Bourgeois2015,bourgeois2016}.

The pulse powers and lengths optimal for readout may not be optimal for initialization and conversion of \nvn~to \nv. Therefore in Fig.~\ref{tion} b) we introduce a second laser pulse to separate the ionization process from the \nv~initialization. Here $\Delta I/I$ is plotted against the reset pulse length $T_\text{reset}$ for different reset pulse powers $P_\text{reset}$. Small reset pulse powers improve $\Delta I/I$ for increasing $T_\text{reset}$. The optimal $P_\text{reset} = \SI{3.1}{\milli\watt}$ leads to a maximal $\Delta I/I$ of \SI{-17}{\percent} which is reached for $T_\text{reset}$ longer than \SI{3}{\micro\second}. For higher $P_\text{reset}$ the reset pulse itself starts to ionize the \nv~centers which leads to a decrease of $\Delta I/I$ since the \nv~have already undergone spin-dependent ionization or a decay through the ISC which eradicates all spin information.  

\begin{figure}[t]
\centering
\includegraphics[width=\linewidth]{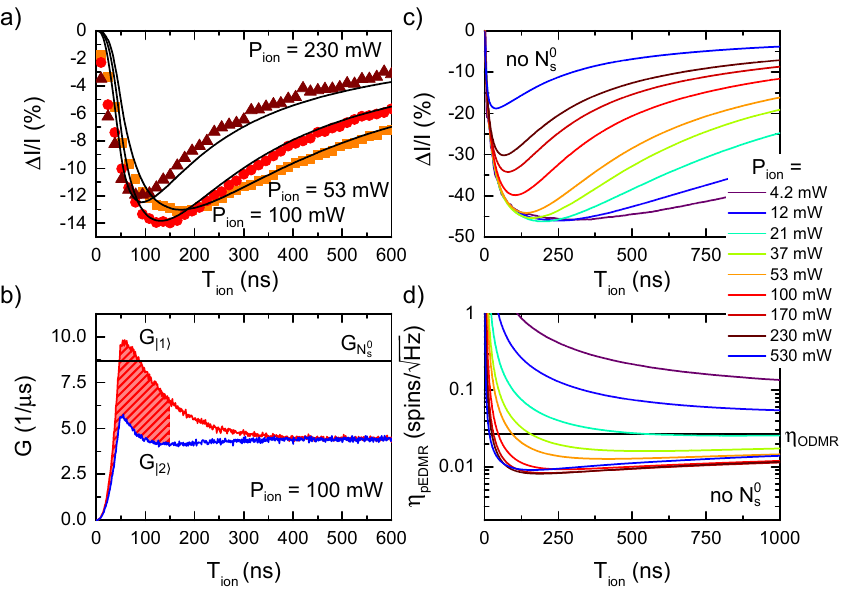}
\caption{a) Contrast $\Delta I/I$ as a function of the length of the ionization pulse $T_\text{ion}$ for three pulse powers taken from Fig.~\ref{tion} and simultaneous fit of a Monte-Carlo simulation (black lines). b) Simulated charge carrier generation rate $G_\ket{1}$ and $G_\ket{2}$ plotted as a function of the time during the ionization pulse $T_\text{ion}$ for starting states $\ket{1}$ and $\ket{2}$, resp. The black line depicts a constant background charge carrier generation rate $G_{\nsnm}$ originating from substitutional nitrogen donors. c) Simulated $\Delta I/I$ as function of $T_\text{ion}$ for different $P_\text{ion}$ under optimal conditions. d) Sensitivity $\eta_\text{pEDMR}$ as a function of $T_\text{ion}$ for different $P_\text{ion}$ under optimal conditions. $\eta_\text{ODMR}$ marks the typical ODMR sensitivity. }
\label{model}
\end{figure}
We can quantitatively reproduce these observations by a Monte-Carlo model of the \nv~center's optical cycle together with photoionization and recharging of the \nvn~(Fig.~\ref{process} c)) using the partial lifetimes of \citein{Robledo2011}. The excitation time $t_\text{ex}$ from the \textsuperscript{3}A\textsubscript{2} ground state of the \nv~ to its \textsuperscript{3}E excited state, the characteristic time $a\cdot t_\text{ex}$ of the ionization process and the lifetime of the ionized state $t_{\text{NV}^0}$ are used as parameters in the simulation. 
The simulation is repeated \num{1} million times for each starting state. Whenever the simulation transitions from \nv~to \nvn~or from \nvn~to \nv, the generation of an electron or a hole in the respective bands at that time is recorded. 
Following the treatment in \citein{Rose63} the photocurrent through the diamond sample is $I=eG\frac{\tau}{T_\text{r}}$, with the elemental charge $e$, the charge carrier generation rate $G$, the typical charge carrier lifetime $\tau$ and the typical transit time $T_\text{r}$ of the charge carriers through the photoconductor device. The ratio of $\tau$ and $T_\text{r}$ is called the photoconductive gain $g=\frac{\tau}{T_\text{r}}$, so that $I=eGg$. 
Since the charge carrier generation rate is not constant throughout the measurement we replace $G$ by its mean $\mean{G}=\frac{1}{T_\text{ion}}\int_0^{T_\text{ion}}G\left(t\right)\dt=\frac{N}{T_\text{ion}}$ with $N$ the number of charge carriers generated during the laser pulse time $T_\text{ion}$. To account for a background current $I_\text{bg}$, originating from the ionization of $\text{N}_\text{s}^0$, we add the generation of electrons with a rate $G_{\nsnm}=a_\text{bg}/ t_\text{ex}$ which allows to parametrize its dependence on the optical power. Microwave pulse imperfections yielding a mixture between $\ket{1}$ and $\ket{2}$ at the start of the experiment are described by a parameter $p$ multiplied with the contrast curve. The contrast then becomes \begin{align}
\frac{\Delta I}{I}=p\frac{I_\ket{2}-I_\ket{1}}{I_\ket{1}+I_\text{bg}}=p\frac{\mean{G}_\ket{2}-\mean{G}_\ket{1}}{\mean{G}_\ket{1}+G_{\nsnm}}=p\frac{N_\ket{2}-N_\ket{1}}{N_\ket{1}+\frac{a_\text{bg}}{t_\text{ex}}T_\text{ion}},
 \end{align}
where the subscript $\ket{1}$ or $\ket{2}$ denotes the respective value for the initial states $\ket{1}$ and $\ket{2}$.

This term is fitted simultaneously to the data presented in Fig.~\ref{tion} for the three laser powers of \SI{53}{\milli\watt}, \SI{100}{\milli\watt} and \SI{230}{\milli\watt} using the Nelder-Mead simplex algorithm. The fit parameters $a$, $a_\text{bg}$, $p$ and $t_{\text{NV}^0}$ are used globally for all fits while $t_\text{ex}=t_\text{ex,fit}\cdot \SI{100}{\milli\watt}/P_\text{ion}$ is scaled according to the laser power $P_\text{ion}$ used in the different experiments. To keep the complexity of the simulation down we use only one $t_{\text{NV}^0}$ for all three fits which overestimates the generated photocurrent for small laser powers and vice versa. Figure \ref{model} a) compares $\Delta I/I$ and the fit of the Monte-Carlo simulation, which are in very good agreement. We find $t_\text{ex,fit}=\SI{22}{\nano\second}$, $a=\num{1.1}$, $a_\text{bg}=\num{0.19}$, $p=\num{0.75}$ and $t_{\text{NV}^0}=\SI{10}{\nano\second}$. A $t_\text{ex}$ in the range of tens of \si{ns} is in agreement with the onset of a saturation in the photocurrent at $P_\text{ion}=\SI{100}{\milli\watt}$ (cf. Fig.~\ref{tion} c)) which we expect to happen at the point where the excitation time from $\ket{1,2}$ to $\ket{3,4}$ reaches the partial lifetime for the states $\ket{3,4}$ to the state $\ket{1,2}$. $p<\num{1}$ is probably caused by the limited pulse fidelity at $B_0=0$, since differently oriented \nv~centers have different Rabi frequencies.

The model and the parameters determined allow us to simulate the charge carrier generation dynamics in our sample during a laser pulse. Figure \ref{model} b) shows the charge carrier generation rate $G_\ket{1}(t)$ and $G_\ket{2}(t)$ for $P_\text{ion}= \SI{100}{\milli\watt}$. The horizontal line depicts a background current originating from \nsn~at $G_{\nsnm}=\SI{8.6}{\per\micro\second}$. For times bigger than \SI{400}{\nano\seconds} all spin-dependent signal is lost and the system is in a steady state with $\widetilde{G}_\ket{1}=\widetilde{G}_\ket{2}=\SI{4.5}{\per\micro\second}$, each at about $1/2$ the charge carrier generation rate originating from \nsn. For pulsed illumination with a pulse length $T_\text{ion} = \SI{150}{\nano\second}$ we find $N_\ket{1}=\num{1.0}$, $N_\ket{2} = \num{0.6}$ and an \nsn~background of \num{1.3} by integration over the curves. Therefore, even at high laser powers, the contrast is limited by \nsn~ionization. 

In order to find the limits of our detection method in samples free of \nsn~background we simulate $\Delta I / I$ without a background photocurrent,  with instantaneous AOM turn-on, and with $p = \num{1}$. Figure \ref{model} c) plots the contrast $\Delta I / I$ simulated under these conditions versus $T_\text{ion}$. Again, the $\Delta I/I$ has an optimal $T_\text{ion}$ for each $P_\text{ion}$. The most notable difference is the maximal contrast of $\Delta I/I=\SI{-46}{\percent}$ which is predicted for a laser power of \SI{21}{\milli\watt}.

However, we expect maximum sensitivity to be obtained for rather different optical pulse conditions. A sensitivity $\eta$ is usually defined by $\eta= \frac{1}{\text{SNR}}\cdot\frac{N}{\sqrt{\Delta f}}$, with $\text{SNR}$ the signal-to-noise ratio, $N$ the number of spins and $\Delta f$ the detection bandwidth of the particular experiment \cite{Boero2003}. For ODMR we find the $\text{SNR}$ using Poissonian statistics where the difference in photoluminescene (PL) counts $\Delta cts=cts_\ket{1}-cts_\ket{2}$ for the different initial states is divided by the shot noise generated by the number of counts $\sqrt{cts_\ket{1}}$ of the bright initial state $\ket{1}$. For pEDMR we use the difference in the current $\Delta I$ divided by the sum of the shot noise generated by the total current $\sqrt{2eI\Delta f}$ \cite{rauschen} and the amplifier input noise $\delta I_\text{amp}\cdot\sqrt{\Delta f}$. With this we find
\begin{align}
\eta_\text{ODMR} &= \frac{\sqrt{N}}{c\sqrt{cts_\text{single}\Delta f}}\quad\text{and}\\
\eta_\text{pEDMR} &= \frac{\sqrt{2eI_\text{single}N}+\delta I_\text{amp}}{c I_\text{single}},
\end{align}
where the subscript single denotes the corresponding value for a single \nv~center and $c$ is the contrast of the corresponding measurement. Figure~\ref{model} d) plots the simulated sensitivity $\eta_\text{pEDMR}$ at a sequence repetition rate of \SI{500}{\kilo\hertz} and for a typical amplifier input noise of \SI{0.2}{fA\per\sqrthz} versus $T_\text{ion}$. For each $P_\text{ion}$ the sensitivity decreases (i.e.~improves) with longer $T_\text{ion}$  because of the increased current $I_\text{single}$ and contrast $c$ for longer ionization pulse times. After reaching an optimal value $\eta_\text{pEDMR}$ increases again since the decrease in contrast $c$ cancels the effects of the higher currents. The expected optimal sensitivity of \SI{0.008}{spins\per\sqrthz} for pEDMR is not reached for the laser power corresponding to the maximum contrast but rather for $P_\text{ion}$ between \SI{170}{\milli\watt} and \SI{230}{\milli\watt} and $T_\text{ion}\approx\SI{200}{\nano\second}$. For comparison, the sensitivity for a typical ODMR experiment on a single \nv~center with a countrate of \SI{100}{\kilo\cts\per\second}, a contrast of \SI{30}{\percent} and an integration time over the fluorescence of \SI{300}{\nano\second} is $\eta_\text{ODMR}=\SI{0.027}{spins\per\sqrthz}$ which is marked by the black horizontal line in Fig.~\ref{model} d).

To put this sensitivity in absolute numbers: Using a \num{100}x objective and \SI{1.2}{\milli\watt} of laser power a cw photocurrent of \SI{32}{\pico\ampere} is generated in our sample. Comparing the PL observed on it and on a reference sample with single \nv~centers allows us to estimate the number of \nv~participating to about \num{130} so that each \nv~contributes a photocurrent of \SI{240}{\femto\ampere}. Simulations under the corresponding power of \SI{30}{\milli\watt} for the \num{5}x objective predict a current of \SI{580}{\femto\ampere} per \nv, so that the photoconductive gain in our samples is $g=0.35<1$ as expected for a metal-semiconductor-metal photodetector based on Schottky contacts. Under the optimized conditions given above (\SI{170}{\milli\watt} optical power for a \num{5}x objective, \SI{500}{\kilo\hertz} repetition rate, $g=\num{0.35}$, $T_\text{ion}=\SI{200}{\nano\second}$, no background current, no pulse imperfections), a single \nv~should exhibit a $\Delta I = \SI{54}{\femto\ampere}$ on a photocurrent of  $I=\SI{190}{\femto\ampere}$, which should be easily measurable. This corresponds to a 'count rate' of \SI{1.2}{\mega\cts\per\second} elementary charges at a spin contrast of \SI{30}{\percent}, numbers which are not reached for photons in conventional detection setups

In summary, using a combination of pulsed photoionization and pulsed spin manipulation, we have demonstrated electrical readout of the coherent control of an ensemble of \nv~centers. With the help of a Monte-Carlo simulation we have improved our understanding of the photoionization dynamics and find that single-spin (multi-shot) detection should be feasible electrically, possibly with a higher sensitivity than optically. These results motivate a range of further studies, in particular into the relative benefits of photoconductors with ohmic or Schottky contacts and into more advanced photoionization schemes using lasers with different photon energies \cite{shields2015, hopper2016, bourgeois2016}, which could lead to higher ionization efficiencies and a better understanding of the dynamics. Furthermore, EDMR based on photoionization should be transferable to other defects and other host materials such as SiC, which might allow even easier integration of electrical spin readout, e.g., with bipolar device structures.  
  
This work was supported financially by Deutsche Forschungsgemeinschaft via FOR 1493 (STU139/11-2), SPP 1601 (BR 1585/8-2) and  Emmy Noether grant RE 3606/1-1.
\bibliography{papernv}

\end{document}